\documentclass[12pt]{article}
\usepackage[final]{graphicx}  

\usepackage{amsmath,amssymb} 
\usepackage[square,numbers]{natbib}
\begin{document}
\title{On Classical Analogs of Quantum Schwarzschild and Reissner-Nordstrom Black Holes. Solving the "Mystery of $\log{3}$"}


\author{Victor Berezin\\
        Institute for Nuclear Research \\
        Russian Academy of Sciences\\
        Moscow, 117312, 60-October Anniversary pr.,7-a\\
        E-mail:berezin@ms2.inr.ac.ru}

\maketitle
\begin{abstract}
The model is built in which the main global properties of classical
and quasi-classical black holes become local. These are the event
horizon, "no-hair", temperature and entropy. Our construction is
based on the features of a quantum collapse, discovered while
studying some quantum black hole models. But it is purely classical,
and this allows to use the Einstein equations and classical (local)
thermodynamics and explain in this way the "$\log{3}$" - puzzle.
\end{abstract}

Classical definition of the black hole is based on the existence of
the event horizon \cite{HE}. The notion of the latter is global and
requires the knowledge of the whole history, both past and future.
Classical "black hole has no hair" \cite{RuW} and is described by
only few parameters. The process of becoming bald is also global,
its duration, formally, is infinite. At late times the frequencies
of the decaying modes are complex with equidistantly growing
imaginary parts, they are called quasi-normal frequencies. Their
real parts approaches the finite limit. The appearance of damping
oscillations of this type points to the existence of some resonance
frequency inherent in a black hole. Investigation of the processes
near the event horizon showed that they can be reversible and
irreversible \cite{Cr,CrR}, like in thermodynamics, and the black
hole area cannot decrease.

These features were generalized by suggestion \cite{Bek} that the
Schwarzschild black hole, indeed, can be considered as some thermal
equilibrium state having both the temperature and entropy, the
latter being a reflection of the "no hair" property, i.e., the black
hole with given parameters could be formed by enormously many
different ways. The suggested proportionality of the black hole
entropy to the horizon area was then confirmed by extending the four
laws of thermodynamics to the general type of black holes,
possessing mass electric charge and angular momentum \cite{BCH}.
Finally, calculations made by S.Hawking \cite{H} showed that the
temperature is real, black holes should evaporate, and the entropy
is one fourth of the horizon area.

Numerous attempts to quantize black holes taught us that their mass
spectrum is discrete, and the temperature and entropy are the
properties of the quasi-classical stage. The detailed description
and understanding of such a quasi-classical regime is very
important, but at the same time is very difficult because of the
global character of the main features of the classical black holes.
In this paper we construct the so-called classical analogs of
quantum black holes for which global properties of classical and
semiclassical black holes become local, what makes their description
and understanding much more easy.

Our starting point is a quantum mechanical model for a spherically
symmetric self-gravitating thin dust shell. Such a shell is a
simplest generalization of a point particle with the advantage that
it has the dynamical degree of freedom - the shell radius, and the
corresponding classical problem has the exact solution with the full
account for back reaction of the matter source on the space-time
metric. Due to the spherical symmetry the shell's radius is the only
dynamical variable of the whole system: space-time plus a
gravitating source, and the quantum functional Wheeler-DeWitt
equation is reduced to the following stationary Schroedinger
equation in finite differences \cite{B,BBN}:
\begin{equation}
\label{Psi}
\Psi(m, m_{in}, S + i\zeta) + \Psi(m, m_{in}, S - i\zeta) =
\frac{F_{in} + F_{out} - \frac{M^2}{4 m^2 S}}{\sqrt{F_{in}} \sqrt{F_{out}}}
\Psi (m, m_{in}, S) ,
\end{equation}
where $m = m_{out} = m_{tot}$ - the total mass of the system,
$m_{in}$ - the Schwarzschild mass inside, $M$ is the bare mass of
the shell, $S = \frac{R^2}{4 G^2 m^2} \,$ ($R$ - radius, $G$ -
gravitational constant), $F = 1 - \frac{2 G m}{R},\, \zeta =
\frac{m_{Pl}^2}{2 m^2} \,$ ($m_{Pl} = \sqrt{\frac{\hbar c}{G}}$ is
the Planckian mass and we use units with $\hbar = c = k = 1$,
$\hbar$ - Planck constant, $c$ - speed of light, $k$ - Bolzmann
constant). By investigation of wave functions in the vicinity of
singular points (infinities and singularities) and around the
branching points (apparent horizons) the following discrete mass
spectrum for bound states was found $(\Delta m = m_{out} - m_{in})$
 \cite{BBN}:
\begin{eqnarray}
\label{spectrum}
\frac{2 (\Delta m)^2 - M^2}{\sqrt{M^2 - (\Delta m)^2}} &=&
\frac{2 m_{Pl}^2}{\Delta m + m_{in}}\, n \,, \nonumber \\
M^2 - (\Delta m)^2 &=& 2 (1 + 2 p)\, m_{Pl}^2 \,,
\end{eqnarray}
where $n$ and $p \ge 0$ are integers. The appearance of two quantum
numbers instead of one in conventional quantum mechanics is due to
the nontrivial causal structure of the complete Schwarzschild
manifold which contains the so-called Einstein-Rosen bridge and,
correspondingly, two isometric regions with spatial infinities.
Unlike the classical shell motion confined to only one of these
regions, the wave function of the quantum shell "feels" both
infinities. The principal quantum number $n$ comes from the boundary
condition at "our" infinity, while the new, second, quantum number
$p$ - from the other one. The above spectrum is not universal in the
sense that the corresponding wave functions form a two-parameter
family $\Psi_{n,p}$. But for the quantum Schwarzschild black hole we
expect a one-parameter family of solutions, because quantum black
holes should not have "no hairs", otherwise there will be no smooth
classical limit. This means that our spectrum is not a quantum black
hole spectrum, and corresponding quantum shells do not collapse
(like an electron in hydrogen atom). Physically, it is quite
understandable, because the radiation was not included into
consideration. Therefore, quantum gravitational collapse (even
spherically symmetric) is accompanied with radiation. This is true
also for the unbound motion because, though a principal quantum
number $n$ disappears in this case, the second quantum number $p$
still exists and the collapsing shell is eventually settled into
some bound state. The appearance of two quantum numbers instead of
one leads to yet another consequence: the quantum gravitational
collapse proceeds via production new shells, increasing the inner
mass $m_{in}$ inside the primary shell. Such a process can go in
many different ways, so, the quantum collapse is accompanied with
the loss of information, thus converting an initially pure quantum
state into some thermal mixed one. But how could quantum collapse be
stopped? The natural limit is the transition from a black hole-like
shell to a wormhole-like shell by crossing an Einstein-Rosen bridge,
since such a transition requires (at least in a quasi-classical
regime) insertion of infinitely large volume, which probability is,
of course, zero. Computer simulations show that the process of
quantum gravitational process stops when the principal quantum
number becomes zero, $n = 0$. The point $n = 0$ in our spectrum is
very special. In this state the shell does not "feel" not only the
outer region (what is natural for the spherically symmetric
configuration), but it does not know anything about what is going on
inside. It "feels" only itself. Such a situation reminds the "no
hair" property of a classical black hole. Finally, when all the
shells (both the primary one and newly produced) are in the
corresponding states $n_i = 0$, the whole system does not "remember"
its own history. And it is this "no- memory" state that can be
called "the quantum black hole". Note, that the total masses of all
the shells obey the relation
\begin{equation}
\label{miMi}
\Delta m_i = \frac{1}{\sqrt{2}} M_i.
\end{equation}
The subsequent quantum Hawking's evaporation can proceed via some
collective excitations.

The final state of quantum gravitational collapse, the quantum black
hole, can be viewed as some stationary matter distribution.
Therefore, we may hope that for massive enough quantum black hole
such a distribution is described approximately by a classical static
spherically symmetric perfect fluid with energy density
$\varepsilon$ and pressure $p$ obeying classical Einstein equations.
This is what we call a classical analog of a quantum black hole. Of
course, in such a case the corresponding classical distribution has
to be very specific. To study its main features let us consider the
situation in more details.

Any static spherically symmetric metric can be written in the form
\begin{equation}
\label{ds}
d s^2 = e^{\nu} d t^2 - e^{\lambda} d r^2 - r^2 (d \vartheta ^2 +
\sin^2{\vartheta} d \varphi ^2).
\end{equation}
Here $r$ is the radius of a sphere with the area $A = 4 \pi r^2, \nu
= \nu (r), \lambda = \lambda (r)$. The Einstein equations are (prime
denotes differentiation in $r$):
\begin{eqnarray}
\label{Ee}
- e^{- \lambda} \left(\frac{1}{r^2} - \frac{\lambda^{\prime}}{r} \right)
+ \frac{1}{r^2} &=& 8 \pi G \varepsilon \,, \nonumber \\
- e^{- \lambda} \left(\frac{1}{r^2} + \frac{\nu ^{\prime}}{r} \right) +
\frac{1}{r^2} &=& - 8 \pi G p \,, \nonumber \\
- \frac{1}{2} \left( \nu^{\prime \prime} + \frac{\nu ^{\prime^2}}{2} +
\frac{\nu ^{\prime} - \lambda ^{\prime}}{r} -
\frac{\nu ^{\prime} \lambda ^{\prime}}{2} \right) &=& - 8 \pi G p \,.
\end{eqnarray}
We see that there are three equations for four unknown functions of
one variable, namely, $\nu(r), \lambda(r),$ $\varepsilon(r)$ and
$p(r)$. But, even we would know an equation of state for our perfect
fluid, $p = p(\varepsilon)$, the closed (formally) system of
equations would have too many solutions. We need, therefore, some
selection rules in order to single out the classical analog of
quantum black hole. Surely, the "no hair" feature should be the main
criterium. Thus, we have to adjust our previous definition of the
"no-memory" state to the case of a continuum matter distribution.
For this, let us integrate the first of Eqns.(\ref{Ee}):
\begin{equation}
\label{F}
e^{- \lambda} = 1 - \frac{2 G m(r)}{r},
\end{equation}
where
\begin{equation}
\label{mr}
m(r) = 4 \pi \int\limits_0^r \varepsilon \tilde r^2 d \tilde r
\end{equation}
is the mass function that must be identified with $m_{in}$. Now, the
"no memory" principle is readily formulated as the requirement, that
$m(r) = a r$, i.e.,
\begin{eqnarray}
\label{constlambda} e ^{- \lambda} &=& 1 - 2 G a = const \, , \nonumber \\
\varepsilon &=& \frac{a}{4 \pi \, G \, r^2}.
\end{eqnarray}
Note, that in static case, the inverse metric coefficient $e^{-
\lambda}$ is an invariant which in the general spherically symmetric
space-time reads as $\Delta = - e^{- \lambda} = g^{ik} R_{, i} R_{,
k}$ and is nothing more but a squared normal vector to the surface
of constant radius $R(x^i) = R(t, q) = const$. We can also introduce
a bare mass function $M(r)$ (the mass of the system inside a sphere
of radius $r$ without the gravitational mass defect).

\begin{equation}
\label{Mr}
M(r) = \int \varepsilon dV = 4 \pi \int\limits_0^r \varepsilon (\tilde r)
 e^{\frac{\lambda}{2}}(\tilde r) \tilde r^2 d \tilde r =
 \frac{a r}{\sqrt{1 - 2 G a}} \,.
\end{equation}
The remaining two equations (\ref{Ee}) can now be solved for $p(r)$
and $e^{\nu}(r)$. The general solution is rather complex, but the
correct non-relativistic limit for the pressure $p(r)$ (we are to
reproduce the famous equation for hydrostatic equilibrium) has only
the following one-parameter family:
\begin{equation}
\label{pres}
p(r) = \frac{b}{4 \pi r^2}\, ,
\end{equation}
where
\begin{equation}
\label{b}
b = \frac{1}{G} \left( 1 - 3 G a - \sqrt{1 - 2 G a} \sqrt{1 - 4 G a} \right) \,.
\end{equation}
We see that the solution exists only for $a \le \frac{1}{4 G}$, then
$b \le a$. The physical meaning of these inequalities is that the
speed of sound cannot exceed the speed of light, $v_{sound}^2 =
\frac{b}{a} \le 1 = c^2$, the equality being reached just for $a = b
= \frac{1}{4 G}$. Finally, for the temporal metric coefficient
$g_{00} = e^{\nu}$ we get:
\begin{equation}
\label{nu}
e^{\nu} = C_0 r^{\frac{4 b}{a + b}} = C_0 r^{2 G \frac{a + b}{1 - 2 G a}} \,.
\end{equation}
Thus, demanding the "no-memory" feature and existence of the correct
non-relativistic limit, we obtained the two-parameter family of
static solutions. But we need a one-parameter family, so we have to
continue our search.

Calculation of the Riemann curvature tensor $R^{\mu}_{\nu \lambda
\sigma}$ shows that it is divergent at $r=0$ for $b<a$. But, if $a =
b = \frac{1}{4 G}$ we are witnessing a miracle, the (before)
divergent components become zero, and the remaining nonzero ones
equal
\begin{eqnarray}
\label{Riemann1}
R^0_{202} &=& - (1 - 2 G a) = - \frac{1}{2}\,, \;\;  \left(R^2_{020} =
\frac{1}{2} C_0^2 \right)\, ; \nonumber \\
R^0_{303} &=& - (1 - 2 G a) = - \frac{1}{2}\,, \;\;  \left( R^0_{030} =
\frac{1}{2} C_0^2 \right)\, ; \nonumber \\
R^2_{323} &=& 2 G a \sin^2 {\theta} = \frac{1}{2} \sin^2{\theta}\,,\;\;
\left( R^3_{232} = \frac{1}{2} \right) \,,
\end{eqnarray}
and the only nonzero component of the Ricci tensor $R_{\mu \nu} ( =
R^{\alpha}_{\mu \alpha \nu})$ equals to
\begin{equation}
\label{Ricci}
R_{00} = C_0^2 .
\end{equation}
Thus, demanding, in addition to the previous two very natural
requirements, the third one (also natural), namely, the absence of
the real singularity at $r =0$, we arrive at the following
one-parameter family solutions to the Einstein equations (\ref{Ee}):
\begin{eqnarray}
\label{soln}
g_{00} &=& e^{\nu} = C_0^2 r^2 , \nonumber \\
g_{11} &=& - e^{\lambda} = - \sqrt{2} , \nonumber \\
\varepsilon &=& p = \frac{1}{16 \pi G r^2} .
\end{eqnarray}
So, the equation of state of our perfect fluid is the stiffest
possible one. The constant of integration $C_0$ can be determined by
matching the interior and exterior metrics at some boundary radius
$r = r_0$. Let us suppose that for $r > r_0$ the space-time is
empty, so, the interior should be matched to the Schwarzschild
metric, labeled by the mass parameter $m$. Of course, to compensate
the jump in the pressure $\Delta p \,( = p(r_0) = p_0)$ we must
include in our model some surface tension $\Sigma$, so, actually, we
are dealing with a some sort of liquid. It is easy to check, that
\begin{eqnarray}
\label{match}
C_0^2 &=& \frac{1}{2 r_0^2}\, ; \;\; \Delta p = \frac{2 \Sigma}{\sqrt{2} r_0} \,;
\nonumber \\
e^{\nu} &=& \frac{1}{2} \left(\frac{r}{r_0} \right)^2 ;\;\; p_0 = \varepsilon _0
= \frac {1}{16 \pi G r_0^2} \,;\nonumber \\
m &=& m_0 = \frac{r_0}{4 G}\, .
\end{eqnarray}
Note, that the bare mass $M = \sqrt{2} m$, the relation is exactly
the same as for the shell "no memory" state (\ref{miMi}), and $r_0 =
4 G m_0$, so, the size of our analog of quantum black hole is twice
as that of classical black hole. But how about the special point in
our solution, $r = 0$? It is not a trivial coordinate singularity,
like in a three-dimensional spherically symmetric case, because
\begin{equation}
\label{ds0}
ds^2 (r = 0) = 0.
\end{equation}
This looks rather like an event horizon. Indeed, it can be easily
shown that the two-dimensional $(t-r)$-part of our metric describes
a locally flat manifold. Since the static observers at $r=const$
are, in fact, accelerated, this is a Rindler space-time with the
event horizon at $r=0$. By definition, the surface $r = 0$ can not
be crossed and it is in this sense that the generally global event
horizon becomes local. The corresponding Rindler parameter which in
more general case is called the "surface gravity $\varkappa$",
equals
\begin{equation}
\label{kappa}
\varkappa = \frac{1}{2} \left|\frac{d \nu}{d r}\right| e^{\frac{\nu - \lambda}{2}}
= \frac{C_0}{\sqrt{2}} = \frac{1}{2 r_0}.
\end{equation}
It is well known that uniformly accelerated Rindler observers
register particles with Planckian spectrum. The corresponding
temperature was calculated by W.G.Unruh by investigating a quantum
field theory on two-dimensional manifolds with an event horizon
\cite{U}. The Unruh temperature equals
\begin{equation}
\label{Unruh}
T_U = \frac{a}{2 \pi}.
\end{equation}
The same value can be obtained by considering an Euclidean version
of the Rindler space-time, demanding the absence of conical
singularity that replaces there the event horizon, and then equating
the inverse period of the imaginary time to the temperature.
Calculated in this way, it can be called the "topological
temperature", $T_{top}$. The absence of the conical singularity
means that the unfolded cone has no angle deficit, and the period of
the corresponding azimuth angle equals $2 \pi$. One must remember
that the temperature is not an invariant (scalar) but the temporal
component of the heat flow four-vector, so, its value depends on the
choice of clocks (= time coordinate). The Unruh temperature $T_U$ is
the temperature measured by the observer who is using the proper
time, i.e., with $g_{00} = 1$. When the same observer is using some
local time, then he must deal with the local temperature
\begin{equation}
\label{Tl}
T_{loc} = \frac{T_U}{\sqrt{g_{00}}}.
\end{equation}

Up to now the model is the same as was elaborated by the author in
2003 \cite{Ber}. But at this point we encounter a dilemma. The
problem is that writing the Unruh temperature in terms of the total
mass $m$ we get
\begin{equation}
\label{Tinf}
T_U = \frac{1}{4 \pi r_0} = \frac{1}{16 \pi \, G \, m}
\,,
\end{equation}
what is two times less than the Hawking temperature \cite{H}
\begin{equation}
\label{TH}
T_H = \frac{1}{8 \pi \, G \, m} \, .
\end{equation}
This is the Unruh temperature measured by the observer sitting at
rest just at the event horizon and, at the same time, by the distant
inertial observer at spatial infinity where $g_{00} = 1$. Adopting
the "natural" boundary condition that the temperature inside is
equal to that of outside we get in the Euclidean section a
discontinuity (the period of the azimuthal angle corresponding to
the imaginary time equals $\pi$ inside and $2 \pi$ outside). In
addition, as one can easily check, we have a jump in the local
temperatures measured by the Rindler observers just inside and
outside the boundary $r= r_0$. Such a jump is only partly
compensated by the surface tension $\Sigma$ and can be considered as
caused the start of the irreversible process of converting the
energy (mass) of the inner region into the radiation. Besides, it is
difficult to explain why local observers just inside and outside the
boundary $r=r_0$ who know nothing about what is going on elsewhere
(especially in the Euclidean section) should give different
interpretations to the intensities of particle creation in heir
detectors. On the other hand, if we make the second choice, i.e.,
that the temperature inside is just the Unruh temperature for the
corresponding Rindler space-time, $T_U = \frac{1}{4 \pi \, r_0}$,
then everything is smooth in the Euclidean region and there is no
problem with local observations, but now we should somehow explain
the origin of the Hawking temperature $T_H = \frac{1}{8 \pi \,G \,
m}$ because the latter is obviously measured at infinity by
detecting the heat flux.

Before coming to the thermodynamics we should describe two
interesting and very important features of quasi-classical black
holes. The first of them is the quantization of entropy. In 1973
J.Bekenstein made the remarkable observation \cite{Bek} that the
horizon area of a non-extremal black hole behaves as a classical
adiabatic invariant. In the spirit of the Ehrenfest principle he
conjectured that the horizon area and, therefore, the quantum black
hole entropy, should have a discrete spectrum of the form
\begin{equation}
\label{BE}
S = \gamma \, n, \;\;\;\; n = 1,2,3,...
\end{equation}
Applying statistical physics arguments, J.Bekenstein and V.Mukhanov
showed \cite{M,BM} that the spacing coefficient must be equal to
$\gamma = \log {k} \, , \;\;k = 2,3,...$. Such a value does not
contradict the $\log{2}$-prediction coming from the information
theory which connects the entropy production to the information
loss, and the very famous claim by J.A.Wheeler "It from Bit". The
second feature is the existence of the proper frequency, inherent in
the black hole with given parameters. This frequency was discovered
when studying the behavior of various types of perturbations
(scalar, vector, tensor) around a black hole (see, e.g., \cite {C})
in the attempts to understand how the process of gravitational
collapse proceeds resulting eventually in the black hole baldness.
The evolution of a small perturbation is governed by a
one-dimensional Schroedinger-like wave equation, first derived by
T.Regge and J.A.Wheeler \cite{RW} in the case of Schwarzschild black
hole. For scalar massless (long range) perturbations it reads as
follows, assuming the time dependence of the form $e^{-i \omega t}$
:
\begin{equation}
\label{SlE}
\frac{d^2 \Psi}{d r^{\star 2}} + \left [ w^2 - V(r) \right] \Psi = 0 \, ,
\end{equation}
where the tortoise radial coordinate $r^{\star}$ is related to the
radius $r$ by $dr^{\star} = \frac{dr}{1 - \frac{2 G m}{r}}, \, m$ is
the Schwarzschild black hole mass, and the effective potential is
given by ($l$ is the multipole moment)
\begin{equation}
\label{scpot}
V(r) = \left ( 1 - \frac{2\, G \, m}{r} \right)
\left (\frac{l(l+1)}{r^2} + \frac{2}{r^3} \right) \,.
\end{equation}
It appeared that, at late times, all perturbations are radiated away
in a manner reminiscent of the last pure dying tones of a ringing
bell. To describe these free oscillations of the black hole the
notion of quasi-normal modes was introduced \cite{P}. The
quasi-normal frequencies (ringing frequencies) are characteristic of
the black hole itself, they correspond to solutions of the above
wave equation with the physical boundary conditions of purely
outgoing waves at spatial infinity ($r^{\star} \to \infty$) and
purely ingoing wave crossing the event horizon ($r^{\star} \to
-\infty $). There are infinite number (for a given harmonic index)
of complex frequencies with decreasing relaxation times, i.e.,
increasing imaginary parts. Their real parts, on the other hand,
approaches an asymptotic constant value. For the Schwarzschild black
hole of mass $m$ the quasi-normal frequencies equal ($n \gg 1$)
\cite{N}
\begin{equation}
\label{QNM}
G \, m \, w_n = 0.0437123 - \frac{i}{4} \left (n + \frac{1}{2}\right ) + O[(n+1)^{-1/2}].
\end{equation}
In 1998 S.Hod recognized \cite{Hod} that the real part is actually
equal to $\frac{\log {3}}{8 \pi}$ and, using the famous Bohr's
corresponding principle: "transitions frequencies at large quantum
numbers should equal classical oscillation frequencies" and the
relations $dm = Re \, w_n ,\, A = 4 \pi r_g^2 = 16 \pi G^2 m^2 = 4
G\, S $, deduced that $\gamma = \log{3}$. This value is also in
agreement with the general result obtained by J.Bekenstein and
V.Mukhanov, but contradicts the value $\log{2}$ advocated by the "It
from Bit" claim.

Note, that both the entropy quantization and the quantum nature of
radiation suggest the discrete nature of the quantum (and,
correspondingly, quasi-classical) black hole constituents. We can
imagine some number of quasi-particles, black hole phonons,
interference between which results in equidistant spectrum of
excitations, and transition from different energy states to their
neighbors produces quanta of quasi-normal frequencies.

Let us proceed with the thermodynamics. We begin with the first
choice for the temperature and, following the line of reasoning
presented in \cite{Ber}, have a look at the result. In what follows
we distinguish between two types of thermodynamic relations, the
local ones as seen and measured by the local static observer, and
the global for the distant inertial observer at infinity. The local
observer deals with the bare mass $M$ defined as the following
integral over some volume $V$:
\begin{equation}
\label{Mint}
M = \int T^{0 \lambda} \xi_{\lambda} dV = \int T^0_0 \xi^0 dV =
\int \varepsilon dV \, ,
\end{equation}
where $T^{\lambda}_{\nu}$ is the energy-momentum tensor, $\xi^{\mu}$
- the Killing vector normalized as $\xi^0 = 1$. Thus, this observer
is using the local time and measures the local temperature $T_{loc}
= \frac{T_U}{\sqrt{g_{00}}} = \frac{1}{\sqrt{2} \pi r}$. The first
law of thermodynamics now reads as follows
\begin{equation}
\label{fth}
dM = \varepsilon dV = T_{loc} dS - p dV + \mu dN \,.
\end{equation}
Here $\mu$ is the chemical potential related to the number of black
hole phonons (this is how the integer number enters our model), it
ought to be included because in our model all the distributions are
universal and the only parameter that changes is the boundary value
of radius $r_0$, and this means the automatical changing of all the
integrated extensive variables, $M, S, V$ and $N$. Dividing the
above expression by the volume element $dV$ we get the first law in
its local form
\begin{equation}
\label{fthl}
\varepsilon (r) = T_{loc}(r) s(r) - p(r) + \mu (r) n(r) \, ,
\end{equation}
where $s$ and $n$ are the entropy and particle densities,
respectively. In our model $\varepsilon = p$, but what about $s$?
The local observer can ask his global counterpart who is educated
enough and knows that the total entropy of the black hole of mass
$m$ is $S = 4 \pi G m^2 = \frac{\pi r_0^2}{4 G}$. Having this
information, our local observer can deduce that
\begin{equation}
\label{ed}
s(r) = \frac{1}{8 \sqrt{2} G r}
\end{equation}
and
\begin{equation}
\label{epsT}
s(r) T(r) = \frac{1}{16 \pi G r^2} \, .
\end{equation}
Remembering now that $\varepsilon = \frac{1}{16 \pi G r^2}$ we
obtain
\begin{equation}
\label{eqall}
\varepsilon(r) = p(r) = s(r) T(R) = \mu (r) n(r) \, .
\end{equation}
Our system is in thermal equilibrium because the Unruh temperature
is constant everywhere in the inner region, and for the local
temperature we have the well known relation $T_{loc} \sqrt{g_{00}} =
const$. But in the thermal equilibrium also $\mu \sqrt{g_{00}} =
const$, hence $\frac{T_{loc}(r)}{\mu (r)} = const$ and,
consequently, $\frac{s(r)}{n(r)} = const$. From this we obtain the
equidistant quantization for the entropy $(\gamma = const, \, N -
integer)$
\begin{equation}
\label{quantentr}
S = \gamma N \, ,\;\;\; N = 1,2,3... \,.
\end{equation}
The spacing coefficient $\gamma$ is universal (does not depend on
the value of $r_0$) and can be calculated by noticing that the free
energy density $f(r)$ is exactly zero
\begin{equation}
\label{fe}
f(r) = \varepsilon(r) - T_{loc}(r) s(r) = 0 \, .
\end{equation}
To do this, let us construct the partition function $Z_1$ for a
small part of our system corresponding to one black hole phonon:
\begin {equation}
\label{pf} Z_1 = \Sigma _n e^{-\frac{\varepsilon_n}{T}} \, .
\end{equation}
Here $\varepsilon_n$ are the excitation energy levels and, since
$\frac{\varepsilon_n}{T}$ is invariant under the change of time
variable (clocks), we will use the proper time of local observers,
so the temperature is just the Unruh temperature, $T = T_U = const$.
What concerns the energy spectrum $\varepsilon_n$, we already
mentioned that the existence of the intrinsic frequency $\omega$ for
Schwarzschild black holes and the equidistant imaginary parts of the
quasi-normal frequencies suggests the following relation (the phonon
spectrum)
\begin{equation}
\label{epsn}
\varepsilon_n = \omega n\, , \;\;\; n = 1,2,3... \, .
\end{equation}
After substituting this into the exponent in Eqn.(\ref{pf}) the
summation can be easily performed:
\begin{equation}
\label{pf1} Z_1 = \frac{e^{-\frac{\omega}{T}}}{1 -
e^{-\frac{\omega}{T}}} \, .
\end{equation}
The transition from the $n'th$ energy state to the $(n+1)'th$ (or
the other way around) gives $dM$, hence $\frac{\omega}{T} = dS_{min}
= \gamma$. From zero value for the free energy we have $Z_1 = 1$,
and
\begin{equation}
\label{l2}
\frac{\omega}{T} = \gamma = \log{2} \, .
\end{equation}

Let us summarize what we have got with the first choice for the
temperature, $T_U = T_H$.

The good features are the equidistant quantization of the entropy
and the value $\log{2}$ for its spacing. And, of course, the very
appearance of the temperature itself, but this is common for given
distributions $\varepsilon(r)$ and $p(r)$ irrespective of the choice
of the temperature $T_U$. From that fact that the inner distribution
is in thermal equilibrium there comes out one more desirable
property which can be called "indifference". If we remove (radiate
away) some outer layer, the inner part would remain unperturbed. The
"indifference" reflects the universality of our classical analog of
quantum black holes which, in turn, is the "analytical continuation"
of the classical black holes. Indeed, the energy density
distribution is universal, the speed of sound equals the speed of
light, the ratio of the resonance frequency to the Unruh temperature
is the universal constant, thus explaining the rather unusual
inverse proportionality of the black hole temperature to its mass.
On the other hand, the inverse proportionality of the resonance
frequency to the mass becomes quite understandable, because it is
translated into inverse proportionality to the boundary radius, in
direct analogy with the music instruments - the smaller the size,
the higher the dominant tone, i.e., the resonance frequency.

The bad features (or, better, "not good") are the following. The
total free energy of the Schwarzschild black hole equals $F - T_H S
= \frac{m}{2}$, and it is impossible to explain why the inner part
of our model has zero free energy. Also, it is impossible to imagine
how to obtain the $\log{3}$ in the real part of the quasi-normal
frequencies.

The crucial test for the validity of our model, i.e., for the choice
of the temperature, is the possibility of its generalization to a
physically acceptable classical analog of the quantum
Reissner-Nordstrom black hole. We constructed two of them
\cite{Talks}. The first model has continuous distributions both of
mass and electric charge, while in the second the charge is
concentrated in the thin massive shell at the boundary surface, the
inner mass distribution being the same as in the Schwarzschild case.
Let us describe briefly the first model. There are two parameters
characterizing it completely, the boundary radius $r_0$ and the
charge/mass ratio $\frac{e^2}{G m^2}$. The radius $r_0$ is again a
free parameter, and everything else: the energy density
$\varepsilon$, radial and tangential pressures $p_r$ and $p_t$,
electric charge distribution $e(r)$ and the surface tension
$\Sigma$, depends solely on the charge/mass ratio. The temperature
matching condition $T_U = T_H$ results in some strange features of
these parametric dependencies. The most awful is the change of sign
in the radial pressure $p_r$ (though energy dominance conditions are
not violated) and, consequently, the change of sign in the surface
tension $\Sigma$, the latter points to the instability - a potential
wall (barrier) is substituted by a potential well. Moreover, the
lack of the "indifference" means that the radiation of a single
quantum with the charge/mass ration different from that of the given
distribution would cause a complete "reloading" of the whole system,
what is unacceptable and nonphysical for large semiclassical black
holes. In the second model with shell-like distribution of the
electric charge such an unpleasant thing is, of course, absent. But,
unfortunately, this model has no smooth limit to the Schwarzschild
uncharged case. Namely, the mare mass of the shell (and, thus, its
total mass) does not vanishes when the charged becomes zero.

Thus, we are forced to make the second choice for the temperature,
\begin{equation}
\label{uh}
T_U = \frac{1}{2} T_H \, ,
\end{equation}
i.e., the Unruh temperature of the inner region is only one half of
the Hawking temperature measured at infinity. This jump is exactly
compensated by the surface tension. Classically, the radiation is
now absent because the heat in the inner region is thermodynamically
locked by this surface tension which provides the equilibrium
temperature gradient from the inner to outer Rindler observers. In
this sense the model is self-consistent since no more back reaction
corrections are needed. In quantum theory the radiation will be
caused by the tunneling process.  But let us proceed with the
thermodynamical relations, and write down the first law of
thermodynamics in the local form,
\begin{equation}
\label{nlth}
\varepsilon(r) = T_{loc}(r) s(r) - p(r) + \mu(r) n(r) \, .
\end{equation}
Again, $\varepsilon(r) = p(r) = \frac{1}{16 \pi G r^2}$, but now
$T_{loc}(r) = \frac{1}{2 \sqrt{2} \pi r}$, and we have
\begin{equation}
\label{nfthl}
T_{loc}(r) s(r) = \frac{1}{2} \varepsilon \, , \;\;\;
\mu(r) n(r) = \frac{3}{2} \varepsilon \, .
\end{equation}
From this it follows that the free energy is no more zero, but
\begin{eqnarray}
\label{nfe}
f(r) &=& T_{loc}(r) s(r) = \frac{1}{2} \varepsilon (r)\, , \nonumber \\
F(r) &=& \frac{1}{2} M = 2 T_{loc}(r_0) S
\end{eqnarray}
as measured by local observers, $T_{loc}(r_0) = \frac{1}{2 \sqrt{2} \pi r_0},
 \, S$ is the total entropy of the system, $M$ is its bare mass. The
 distant observer at infinity measures the total mass $m =
 \frac{1}{\sqrt{2}} M$ and the Hawking temperature $T_H = 2 T_U =
 \frac{2}{\sqrt{2}} T_{loc} (r_0)$, hence, for him our free energy
 becomes
 \begin{equation}
 \label{finf}
 F_{\infty} = m - T_H S = \frac{1}{2} m \, ,
 \end{equation}
 and this guaranties the usual Schwarzschild black hole
 thermodynamical relation $dm = T_H dS$.

 Surely, we again have the equidistant entropy quantization $S =
 \gamma N$, but with, perhaps, different spacing $\gamma$. To
 evaluate this spacing we need to know the partition function. Of
 course, it is the same as before, i.e., for the one-phonon patch we
 have
 \begin{equation}
 \label{npf}
 Z_1 = \frac{e^{- \frac{\omega}{T}}}{1 - e^{- \frac{\omega}{T}}} \, ,
 \end{equation}
 where, as was explained earlier, $\omega$ is an intrinsic black
 hole proper (resonance) frequency, and $T = T_U $. The total
 partition function equals $Z_{tot} = \left ( Z_1 \right) ^N $. The
 partition function is an invariant. For any small part of our
 system we should have the usual relation between densities of its
 free energy and partition function, $f dV = - T_{loc}
 \log{Z_{small}}$. Integration over the volume gives us
 \begin{equation}
 \label{zint}
 \int \frac{f}{T_{loc}} dV = - \Sigma \log{Z_{small}} dV = - \log{Z_{tot}} \, .
 \end{equation}
 The left hand side equals
 \begin{equation}
 \label{lhs}
 \int \frac{f}{T_{loc}} dV = \frac{1}{2} \int \frac{\varepsilon}{T_{loc}} dV =
 2 \sqrt{2} \pi \int^{r_0}_0 \frac{\varepsilon}{T_{loc}} r^2 dr =
 \frac{\pi r_0^2}{4 G} = \frac{\pi r_g^2}{G} = S \, .
 \end{equation}
 Here $r_g$ is the Schwarzschild radius, and $S$ is the total black
 hole entropy. Eventually, we obtain the the important
 relation
 \begin{equation}
 \label{stot}
 e^{-S} = Z_{tot} = \left ( Z_1 \right )^N \, ,
 \end{equation}
 from which it follows that
 \begin{eqnarray}
 \label{gammas}
 \frac{e^{- \frac{\omega}{T}}}{1 - e^{-\frac{\omega}{T}}} &=& e^{- \frac{S}{N}}
 = e^{- \gamma} \, , \nonumber \\
 e^{\gamma} &=& e^{\frac{\omega}{T}} - 1 \,.
 \end{eqnarray}
 To go further, let us consider the irreversible process of
 converting the mass (energy) of the system into radiation from a
 thermodynamical point of view. In our model such a process takes
 place just at the boundary $r = r_0$, and the thin shell with zero
 surface energy density and surface tension $\Sigma$ serves as a
 convertor supplying the radiation with extra energy and extra
 entropy, and this resembles the "brick wall" model \cite{tH}. One
 can imagine that the near-boundary layer of thickness $\Delta r_0$
 is converting into radiation, thus decreasing the boundary of the
 inner region to $(r_0 - \Delta r_0)$. Its energy equals $\Delta M =
 \epsilon \Delta V$ plus the energy released from the work done by
 the surface tension due to its shift, which is equal exactly to
 $\Sigma d(4 \pi r_0^2) = p \Delta V = \varepsilon \Delta V = \Delta
 M $. Therefore, both the energy  and its temperature becomes
 two times higher that that for any inner layer of the same
 thickness. And this double energy is gained by the radiating
 quanta. Clearly, they have the double frequency and exhibit double
 temperature, so
 \begin{equation}
 \label{ln}
 \frac{Re\, w}{T_H} = \frac{\omega}{T_U} = \log{3} \, .
 \end{equation}
 Substituting this into Eqn.(\ref{gammas}) we obtain
 \begin{equation}
 \label{fg}
 \frac{e^{- \log{3}}}{1 - e^{- \log{3}}} = \frac{\frac{1}{3}}
 {1 - \frac{1}{3}} = \frac{1}{2} = e^{- \gamma}
 \end{equation}
 and
 \begin{equation}
 \label{ltwo}
 \gamma = \log{2} \, .
\end{equation}
(Note, that substituting in Eqn.(\ref{gammas}) the general value
$\gamma = \log{k}, k = 2,3...$,  we obtain $\frac{\omega}{T} =
\log{(k+1)}$). Since the radiated energy is thermalized, its energy
density decreases during expansion due to the work done by the
pressure and, thus, the interpretation of $dm$ as equal to $Re \,w$
is an improper procedure. This resolves the "$\log{3}$-paradox".

The author would like to thank Alexey Smirnov for numerous helpful
discussions. I am greatly indebted to my wife Anastasia Kouprianova
and to D.O., D.O. and N.O.Ivanovs for the permanent and strong moral
support.

\end{document}